\documentclass[letterpaper, 10 pt, conference]{ieeeconf}  

\IEEEoverridecommandlockouts                              

\usepackage{graphicx}
\usepackage{url}  

\usepackage{amsmath}

\title{\LARGE \bf
Deep Learning Frameworks Applied For \\ Audio-Visual Scene Classification
}

\author{Lam~Pham, 
        Alexander~Schindler,
        Mina~Schütz,
        Jasmin~Lampert,
        Sven~Schlarb,
        Ross~King \\ \\
Competence Unit Data Science \& Artificial Intelligence, \\ Center for Digital Safety \& Security, \\ Austrian Institute of Technology, Austria. \\        
}

\begin{document}

\maketitle
\thispagestyle{empty}
\pagestyle{empty}

\begin{abstract}

In this paper, we present deep learning frameworks for audio-visual scene classification (SC) and indicate how individual visual and audio features as well as their combination affect SC performance.
Our extensive experiments, which are conducted on DCASE (IEEE AASP Challenge on Detection and Classification of Acoustic Scenes and Events) Task 1B development dataset, achieve the best classification accuracy of 82.2\%, 91.1\%, and 93.9\% with audio input only, visual input only, and both audio-visual input, respectively.
The highest classification accuracy of 93.9\%, obtained from an ensemble of audio-based and visual-based frameworks, shows an improvement of 16.5\% compared with DCASE baseline.
\newline


\end{abstract}

\section{Introduction}

Analysing both audio and visual (or image) information from videos has opened a variety of real-life applications such as detecting the sources of sound in videos~\cite{cite_01}, lip-reading by using audio-visual alignment~\cite{cite_02}, or source separation~\cite{cite_03}.
Joined audio-visual analysis shows to be effective compared to the visual only data proven in tasks of video classification~\cite{cite_04}, multi-view face recognition~\cite{cite_05}, emotion recognition~\cite{cite_06}, or video recognition~\cite{cite_07}.
Although a number of audio-visual datasets exist, they mainly focus on human for specific tasks such as detecting human activity~\cite{cite_10}, action recognition~\cite{cite_11}, classifying sport types~\cite{cite_12, cite_09}, or emotion detection~\cite{cite_06}.
Recently, the DCASE community~\cite{dcase_web} has released an audio-visual dataset used for DCASE 2021 Task 1B challenge of classifying ten different scene contexts~\cite{ds_2021_1b}.
We therefore evaluate this dataset by leveraging deep learning techniques, then present main contributions following: 1) We evaluate various deep learning frameworks for audio-visual scene classification (SC), indicate individual roles of visual and audio features as well as their combination within SC task. 2) We then propose an ensemble of audio-based and visual-based frameworks, which help to achieve competitive results compared with DCASE baseline; and 3) We evaluate whether the ensemble proposed is effective for detecting scene contexts early.

The paper is organized as follows: Section 2 presents deep learning frameworks proposed for separate audio and visual data input. Section 3 introduces the evaluation setup; where the proposed experimental setting, metric, and implementation of deep learning frameworks are presented. Next, Section 4 presents and analyses the experimental results. Finally, Section 5 presents conclusion and future work.

\section{Deep learning frameworks proposed}
As we aim to evaluate individual roles of audio and visual features within SC task, deep learning frameworks using either audio or visual input are presented in separate sections. 

\subsection{Audio-based deep learning frameworks}
\begin{table}[t]
    \caption{The VGG14 network architecture used for audio-spectrogram based frameworks (input patch of $128{\times}128{\times}6$)} 
        	\vspace{-0.2cm}
    \centering
    \scalebox{0.95}{
    \begin{tabular}{l c} 
        \hline 
            \textbf{Network architecture}   &  \textbf{Output}  \\
        \hline 
         BN - Conv [$3{\times}3] $@$  64$ -  ReLU - BN - Dr (25\%) & $128{\times}128{\times}64$\\
         BN - Conv [$3{\times}3] $@$  64$ - ReLU - BN - AP - Dr (25\%) & $64{\times}64{\times}64$\\
         
         BN - Conv [$3{\times}3] $@$ 128$ - ReLU - BN - Dr (30\%) & $64{\times}64{\times}128$\\
         BN - Conv [$3{\times}3] $@$ 128$ - ReLU - BN - AP - Dr (30\%) & $32{\times}32{\times}128$\\
         
         BN - Conv [$3{\times}3] $@$ 256$ - ReLU - BN - Dr (35\%) & $32{\times}32{\times}256$\\
         BN - Conv [$3{\times}3] $@$ 256$ - ReLU - BN - Dr (35\%) & $32{\times}32{\times}256$\\
         BN - Conv [$3{\times}3] $@$ 256$ - ReLU - BN - Dr (35\%) & $32{\times}32{\times}256$\\
         BN - Conv [$3{\times}3] $@$ 256$ - ReLU - BN - AP - Dr (35\%) & $16{\times}16{\times}256$\\
         
         BN - Conv [$3{\times}3] $@$ 512$ - ReLU - BN - Dr (35\%) & $16{\times}16{\times}512$\\
         BN - Conv [$3{\times}3] $@$ 512$ - ReLU - BN - Dr (35\%) & $16{\times}16{\times}512$\\
         BN - Conv [$3{\times}3] $@$ 512$ - ReLU - BN - Dr (35\%) & $16{\times}16{\times}512$\\
         BN - Conv [$3{\times}3] $@$ 512$ - ReLU - BN - GAP - Dr (35\%) & $512$\\
                  
         FC - ReLU - Dr (40\%)&  $1024$       \\
         FC - Softmax  &  $C=10$       \\
       \hline 
    \end{tabular}
    }
    \label{table:VGG} 
\end{table}
In audio-based deep learning frameworks, audio recordings are firstly transformed into spectrograms where both temporal and frequency features are presented, referred to as front-end low-level feature extraction.
As using an ensemble of either different spectrogram inputs~\cite{lam01, lam02, lam03, lam04} or different deep neural networks~\cite{lam05, truc_dca_18} has been a rule of thumb to improve audio-based SC performance, we therefore propose two approaches for back-end classification, referred to as audio-spectrogram and audio-embedding frameworks.

The audio-spectrogram approach uses three spectrogram transformation methods: Mel filter (MEL)~\cite{librosa_tool}, Gammatone filter (GAM)~\cite{aud_tool}, and Constant Q Transform (CQT)~\cite{librosa_tool}.
To make sure the three types of spectrograms have the same dimensions, the same setting parameters are used with the filter number, window size and hop size set to 128, 80 ms, 14 ms, respectively.
As we have two channels for each audio recording and apply deltas, delta-deltas on individual spectrogram, we finally generate spectrograms of $128\times704\times6$.
These spectrograms are then split into ten 50\%-overlapping patches of $128\times128\times6$, each which represents for a 1-second audio segment.
To enforce back-end classifiers, mixup data augmentation~\cite{mixup1, mixup2} is applied on these patches of spectrogram before feeding them into a VGGish network for classification as shown in Table~\ref{table:VGG}. 
As shown in Table~\ref{table:VGG}, the VGGish network architecture contains sub-blocks which perform convolution (Conv), batch normalization (BN)~\cite{batchnorm}, rectified linear units (ReLU)~\cite{relu}, average pooling (AV), global average pooling (GAP), dropout (Dr)~\cite{dropout}, fully-connected (FC) and Softmax layers. 
The dimension of Softmax layer is set to $C=10$ which corresponds to the number of scene context classified.
In total, we have 12 convolutional layers and 2 fully-connected layers containing trainable parameters that makes the proposed network architecture like VGG14~\cite{vgg_net}. 
We refer to three audio-spectrogram based frameworks proposed as \textit{audio-CQT-Vgg14, audio-GAM-Vgg14}, and \textit{audio-MEL-Vgg14}, respectively.

In the audio-embedding approach, only the Mel filter is used for generating the MEL spectrogram.
the Mel spectrograms are fed into pre-trained models proposed in~\cite{kong_pretrain} for extracting audio embeddings (i.e. the audio embedding, likely vector, is the output of the global pooling layer in pre-trained models proposed in ~\cite{kong_pretrain}).
In this paper, we select five pre-trained models, which achieved high performance in~\cite{kong_pretrain}, as shown in Table~\ref{table:pre_train_aud}, for evaluating the audio-embedding approach.
As these five pre-trained models are trained on AudioSet~\cite{audioset}, a large-scale audio dataset provided by Google for the task of acoustic event detection (AED), using audio embeddings extracted from these models aims to evaluate whether information of sound events detected and condensed in audio embeddings may be effective for SC task. 
Finally, the audio embeddings are fed into a MLP-based network architecture, as shown in Table~\ref{table:MLP}, for classifying into 10 different scene categories.
We refer to five audio-embedding based frameworks proposed as \textit{audio-emb-CNN14, audio-emb-MobileNetV1, audio-emb-Res1dNet30, audio-emb-Resnet38,} and  \textit{audio-emb-Wavegram}, respectively.

In both approaches, the final classification accuracy is obtained by applying late fusion of individual frameworks (i.e. an ensemble of three predicted probabilities from \textit{audio-CQT-Vgg14, audio-GAM-Vgg14, audio-MEL-Vgg14}, or an ensemble of five predicted probabilities from \textit{audio-emb-CNN14, audio-emb-MobileNetV1, audio-emb-Res1dNet30, audio-emb-Resnet38, audio-emb-Wavegram}).
\begin{table}[t]
    \caption{The MLP-based network architecture proposed for classifying audio/visual embeddings} 
        	\vspace{-0.2cm}
    \centering
    \scalebox{0.95}{
    \begin{tabular}{l c} 
        \hline 
            \textbf{Network architecture}   &  \textbf{Output}  \\
        \hline 
         FC - ReLU - Dr (40\%)&  $8192$       \\
         FC - ReLU - Dr (40\%)&  $8192$       \\
         FC - ReLU - Dr (40\%)&  $1024$       \\
         FC - Softmax  &  $C=10$       \\
       \hline 
           	\vspace{-0.7cm}
    \end{tabular}
    }
    \label{table:MLP} 
\end{table}
\begin{table}[t]
    \caption{The pre-trained models in~\cite{kong_pretrain} proposed for \\ extracting audio embeddings} 
        	\vspace{-0.2cm}
    \centering
    \scalebox{0.95}{
    \begin{tabular}{l c} 
        \hline 
            \textbf{Pre-trained models}   &  \textbf{Embedding dimension}  \\
        \hline 
         1/ CNN14 & $2048$\\
         2/ MobileNetV1 & $1024$\\
         3/ Res1dNet30 & $2048$\\
         4/ Resnet38 & $2048$\\
         5/ Wavegram & $2048$\\
        \hline 

    \end{tabular}
    }
    \label{table:pre_train_aud} 
\end{table}
\begin{table}[t]
    \caption{The network architectures~\cite{keras_app} proposed for directly training image frames or extracting image embeddings} 
        	\vspace{-0.2cm}
    \centering
    \scalebox{0.95}{
    \begin{tabular}{l c c} 
        \hline 
            \textbf{Network architectures}   &  \textbf{Size of image inputs} & \textbf{Embedding dimension}  \\
        \hline 
         1/ Xception &$299{\times}299{\times}3$ & $2048$\\
         2/ Vgg19 &$224{\times}224{\times}3$ & $4096$\\
         3/ Resnet50 &$224{\times}224{\times}3$ & $2048$\\
         4/ InceptionV3 &$299{\times}299{\times}3$ & $2048$\\
         5/ MobileNetV2 &$224{\times}224{\times}3$ & $1280$\\
         6/ DenseNet121 &$299{\times}299{\times}3$ & $1024$\\
         7/ NASNetLarge &$331{\times}331{\times}3$ & $4032$\\
        \hline 
    \end{tabular}
    }
    \label{table:image} 
\end{table}
\subsection{Visual-based deep learning frameworks}
Similar to the audio-based frameworks mentioned above, we also propose two approaches for analysing how visual features affect the SC performance: A visual-image approach where classifying process is directly conducted on the image frame inputs, and a visual-embedding approach where the classification is conducted on image embeddings extracted from pre-trained models.
In both approaches proposed, we use the same network architectures from Keras application library~\cite{keras_app}, which are considered as benchmarks for evaluating ImageNet dataset~\cite{Imagenet} as shown in Table~\ref{table:image}. 
In order to directly train image frame inputs with the network architectures in Table~\ref{table:image}, we reduce the $C$ dimensions of the final fully connected layer ($C=1000$ that equals to the number of object detection defined in ImageNet dataset) to $C=10$ that matches the number of scene categories classified.
The visual-image frameworks proposed are referred to as \textit{visual-image-Xception, visual-image-Vgg19, visual-image-Resnet50, visual-image-InceptionV3, visual-image-MobileNetV2, visual-image-DenseNet121}, and \textit{visual-image-NASNetLarge}, respectively.

Regarding the visual-embedding approach, the network architectures mentioned in Table~\ref{table:image} are trained with the ImageNet dataset~\cite{Imagenet}. 
Then, image frames of the scene dataset are fed into these pre-trained models to extract image embeddings (i.e. the image embedding, likely vector, is the output of the second-to-last fully connected layer of pre-trained models). 
Finally, the extracted image embeddings are fed into a MLP-based network architecture as shown in Table~\ref{table:MLP} for classifying into ten scene categories (Note that we use the same MLP-based network architecture for classifying audio or image embeddings).
The visual-embedding frameworks proposed are referred to as \textit{visual-emb-Xception, visual-emb-Vgg19, visual-emb-Resnet50, visual-emb-InceptionV3, visual-emb-MobileNetV2, visual-emb-DenseNet121,} and \textit{visual-emb-NASNetLarge}, respectively.

Similar to the audio-based approaches, the final classification accuracy of visual-based frameworks is obtained by applying late fusion of individual frameworks (i.e. an ensemble of seven predicted probabilities from seven visual-image based frameworks, or an ensemble of seven predicted probabilities from seven visual-embedding based frameworks).

\section{Evaluation Setting}
\subsection{TAU Urban Audio-Visual Scenes 2021 dataset~\cite{ds_2021_1b}}

This dataset is referred to as DCASE Task 1B Development, which was proposed for DCASE 2021 challenge~\cite{dcase_web}. 
The dataset is slightly unbalanced and contains both acoustic and visual information, recorded from 12 large European cities: Amsterdam, Barcelona, Helsinki, Lisbon, London, Lyon, Madrid, Milan, Prague, Paris, Stockholm, and Vienna. It consists of 10 scene classes: airport, shopping mall (indoor), metro station (underground), pedestrian street, public square, street (traffic), traveling by tram, bus and metro (underground), and urban park, which can be categorised into three meta-class of indoor, outdoor, and transportation.
To evaluate, we follow the DCASE 2021 Task 1B challenge~\cite{dcase_web}, separate this dataset into training (Train.) subset for the training process and evaluation (Eval.) subset for the evaluating process as shown in Table~\ref{table:dataset}.

\subsection{Deep learning framework implementation}
\begin{table}[t]
	\caption{The number of 10-second audio-visual scene recordings corresponding to each scene categories in the Train. and Eval. subsets separated from the DCASE 2021 Task 1B development dataset~\cite{ds_2021_1b}.} 
	\vspace{-0.1cm}
	\centering
	\begin{tabular}{l c c c c} 
		\hline 
		\textbf{Category}         & \textbf{Train.}        & \textbf{Eval.}     \\ 
		\hline 
		Airport 	       & $697$                   & $281$           \\        
		Bus     	       & $806$                   & $327$           \\        
		Metro	 	       & $893$                   & $386$           \\        
		Metro Stattion 	       & $893$                   & $386$        \\        
		Park                   & $1006$                   & $386$        \\        
		Public square          & $982$                   & $387$        \\        
		Shopping mall 	       & $841$                   & $387$       \\        
		Street pedestrian      & $968$                   & $421$          \\        
		Street traffic 	       & $985$                   & $402$         \\        
		Tram 	               & $763$                   & $308$          \\   \hline 
         Total              &  $8646$  & $3645$  \\
                            &  ($\approx$24 hours) & ($\approx$10 hours) \\
		\hline 
	\end{tabular}    
	\label{table:dataset} 
\end{table}
\textbf{Extract audio/visual embeddings from pre-trained models}: Since the pre-trained models, which are used for extracting audio embeddings from~\cite{kong_pretrain}, are built on Pytorch framework, the process of extracting embedding from these models is also implemented with Pytorch framework. 
Meanwhile, we use the Tensorflow framework for extracting visual embeddings as the pre-trained models are built with the Keras library~\cite{keras_app} using back-end Tensorflow.

\textbf{Classification models for audio/visual data}: We use Tensorflow framework to build all classification models in this papers (i.e. Vgg14 and MLP-base network architectures mentioned in Table~\ref{table:VGG} and Table~\ref{table:MLP}, respectively).
As we apply mixup data augmentation~\cite{mixup1, mixup2} for both high-level feature of audio/visual embeddings and low-level feature of audio spectrograms/image frames to enforce back-end classifiers, the labels of the mixup data input are no longer one-hot.
We therefore train back-end classifiers with Kullback-Leibler (KL) divergence loss~\cite{kl_loss} rather than the standard cross-entropy loss over all $N$ mixup training samples:
\begin{align}
    \label{eq:loss_func}
    LOSS_{KL}(\Theta) = \sum_{n=1}^{N}\mathbf{y}_{n}\log(\frac{\mathbf{y}_{n}}{\mathbf{\hat{y}}_{n}})  +  \frac{\lambda}{2}||\Theta||_{2}^{2},
\end{align}
where $\Theta$ denotes the trainable network parameters and $\lambda$ denotes the $\ell_2$-norm regularization coefficient. $\mathbf{y_{c}}$ and $\mathbf{\hat{y}_{c}}$  denote the ground-truth and the network output, respectively. The training is carried out for 100 epochs using Adam ~\cite{Adam} for optimization.

\subsection{Metric for evaluation}
Regarding the evaluation metric used in this paper, we follow DCASE 2021 challenge. 
Let us consider $C$ as the number of audio/visual test samples which are correctly classified, and the total number of audio/visual test samples is $T$, the classification accuracy (Acc. (\%)) is the \% ratio of $C$ to $T$.

\subsection{Late fusion strategy for multiple predicted probabilities}
As back-end classifiers work on patches of spectrograms or image frames, the predicted probability of an entire spectrogram or all image frames of a video recording is computed by averaging of all images or patches' predicted probabilities.
Let us consider $\mathbf{P^{n}} = (\mathbf{p_{1}^{n}, p_{2}^{n},..., p_{C}^{n}})$,  with $C$ being the category number and the \(n^{th}\) out of \(N\) image frames or patches of spectrogram fed into a learning model, as the probability of a test instance, then the average classification probability is denoted as  \(\mathbf{\bar{p}} = (\bar{p}_{1}, \bar{p}_{2}, ..., \bar{p}_{C})\) where,

\begin{equation}
    \label{eq:mean_stratergy_patch}
    \bar{p}_{c} = \frac{1}{N}\sum_{n=1}^{N}p_{c}^{n}  ~~~  for  ~~ 1 \leq n \leq N 
\end{equation}
and the predicted label \(\hat{y}\) for an entire spectrogram or all image frames evaluated is determined as:

\begin{equation}
    \label{eq:label_determine}
    \hat{y} = arg max (\bar{p}_{1}, \bar{p}_{2}, ...,\bar{p}_{C} )
\end{equation}

To evaluate ensembles of multiple predicted probabilities obtained from different frameworks, we proposed three late fusion schemes, namely MEAN, PROD, and MAX fusions.
In particular, we conduct experiments over individual frameworks, thus obtain the predicted probability of each framework as  \(\mathbf{\bar{p_{s}}}= (\bar{p}_{s1}, \bar{p}_{s2}, ..., \bar{p}_{sC})\) where $C$ is the category number and the \(s^{th}\) out of \(S\) framework evaluated. 
Next, the predicted probability after late MEAN fusion \(\mathbf{p_{f-mean}} = (\bar{p}_{1}, \bar{p}_{2}, ..., \bar{p}_{C}) \) is obtained by:

\begin{equation}
    \label{eq:mix_up_x1}
     \bar{p_{c}} = \frac{1}{S} \sum_{s=1}^{S} \bar{p}_{sc} ~~~  for  ~~ 1 \leq s \leq S 
\end{equation}

The PROD strategy \(\mathbf{p_{f-prod}} = (\bar{p}_{1}, \bar{p}_{2}, ..., \bar{p}_{C}) \) is obtained by,
\begin{equation}
\label{eq:mix_up_x1}
\bar{p_{c}} = \frac{1}{S} \prod_{s=1}^{S} \bar{p}_{sc} ~~~  for  ~~ 1 \leq s \leq S 
\end{equation}
and the MAX strategy \(\mathbf{p_{f-max}} = (\bar{p}_{1}, \bar{p}_{2}, ..., \bar{p}_{C}) \) is obtained by,
\begin{equation}
\label{eq:mix_up_x1}
\bar{p_{c}} = max(\bar{p}_{1c}, \bar{p}_{2c}, ..., \bar{p}_{Sc}) 
\end{equation}
Finally, the predicted label  \(\hat{y}\) is determined by (\ref{eq:label_determine}):
\label{ssec:ensemble}
\begin{table}[t]
	\caption{Performance comparison of audio-based frameworks} 
	\vspace{-0.1cm}
	\centering
	\begin{tabular}{l c l c } 
		\hline 
		\textbf{Audio-spectrogram}         & \textbf{Acc.}        & \textbf{Audio-embedding}    & \textbf{Acc.}   \\ 
		\textbf{based models}         &        & \textbf{based models}    &   \\ 
		\hline 
		audio-CQT-Vgg14 	       & 68.3                   & audio-emb-CNN14 &   64.4         \\        
		audio-GAM-Vgg14     	       & 69.6                   & audio-emb-MobileNetV1 &  57.8        \\        
		audio-MEL-Vgg14	 	       & 72.2                   & audio-emb-Res1dNet30 &  58.0         \\        
			       &                    & audio-emb-Resnet38 &  62.7     \\        
		                   &                 & audio-emb-Wavegram &  63.4    \\   
  		\hline 

		MAX Fusion          & 78.0                   & MAX Fusion    &  64.9  \\        
		MEAN Fusion 	       & 79.7                   & MEAN Fusion  &  \textbf{69.6}  \\        
		PROD Fusion      & \textbf{80.4}                   & PROD Fusion &   68.4  \\        
		\hline 
	\end{tabular}    
	\label{table:au_res} 
\end{table}
\begin{figure}[t]
    \centering
    \includegraphics[width =1.0\linewidth]{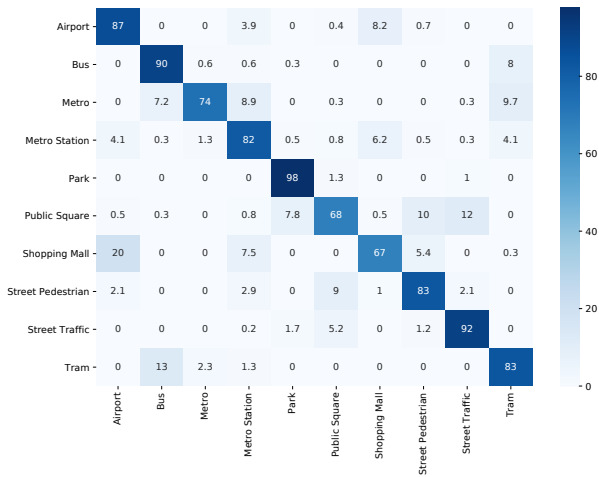}
    	\vspace{-0.5cm}
	\caption{Confusion matrix result (Acc. \%) obtained by PROD fusion of \textit{audio-CQT-Vgg14, audio-GAM-Vgg14, audio-MEL-Vgg14,} and \textit{audio-emb-CNN14}}
    \label{fig:F1}
\end{figure}
\begin{table}[t]
    \caption{Performance comparison of visual-based frameworks} 
        	\vspace{-0.2cm}
    \centering
    \scalebox{0.95}{
    \begin{tabular}{l c l c} 
        \hline 
        \textbf{Visual-image}         & \textbf{Acc.}        & \textbf{Visual-embedding}    & \textbf{Acc.}   \\ 
		\textbf{based models}         &        & \textbf{based models}    &   \\ 
		
        \hline 
         visual-image-Xception &85.9 & visual-emb-Xception& 80.3\\
         visual-image-Vgg19 &83.8 &visual-emb-Vgg19& 80.8\\
         visual-image-Resnet50 &86.3 &visual-emb-Resnet50& 82.0\\
         visual-image-InceptionV3 &88.9 &visual-emb-InceptionV3& 83.4\\
         visual-image-MobileNetV2 &84.4 &visual-emb-MobileNetV2& 80.2\\
         visual-image-DenseNet121 &87.8 &visual-emb-DenseNet121& 83.5\\
         visual-image-NASNetLarge &86.9 &visual-emb-NASNetLarge& 81.5\\
        \hline 
        MAX Fusion          & 90.2                &MAX Fusion     &  \textbf{86.5}   \\        
		MEAN Fusion 	       & 90.5             &MEAN Fusion      &  81.8   \\        
		PROD Fusion      & \textbf{91.1}          &PROD Fusion            &  84.3   \\    
		        \hline 

    \end{tabular}
    }
    \label{table:vi_res} 
\end{table}
\begin{figure}[t]
    \centering
    \includegraphics[width =1.0\linewidth]{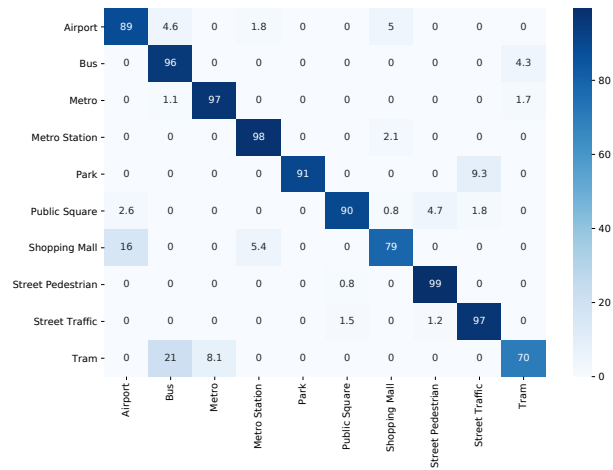}
    	\vspace{-0.5cm}
	\caption{Confusion matrix result (Acc. \%) obtained by PROD fusion of seven visual-image based frameworks}
    \label{fig:F2}
\end{figure}

\section{Experimental Results And Discussion}
\subsection{Analysis of audio-based deep learning frameworks for scene classification}
\label{au_res}
As Table~\ref{table:au_res} shows the accuracy results obtained from audio-based deep learning frameworks, we can see that all late fusion methods help to improve the performance significantly regarding both audio-spectrogram and audio-embedding approaches, achieving the highest score of 80.4\% from PROD fusion of three audio-spectrogram based frameworks and 69.6\% from MEAN fusion of five audio-embedding based frameworks.
Compare the performance between two audio-based approaches proposed, it can be seen that directly training spectrogram inputs is more effective, achieving 68.3\%, 69.6\%, and 72.2\% from CQT, GAM, and MEL spectrogram respectively, which outperform all results obtained from audio-emmbedding based frameworks.
We further conduct PROD fusion of predicted probabilities obtained from three audio-spectrogram based frameworks (\textit{audio-CQT-Vgg14, audio-GAM-Vgg14, audio-MEL-Vgg14}) and the \textit{audio-emb-CNN14} framework (the best framework in the audio-embedding based approach), achieving the classification accuracy of 82.2\% with the confusion matrix shown in Fig.~\ref{fig:F1} and improving the DCASE baseline by 17.1\% (Note that only audio data input is used for these frameworks and DCASE baseline).
This proves that although the audio-embedding based approach shows low performance rather than the audio-spectrogram based approach, audio embeddings extracted from AudioSet dataset for AED task contain distinct features which is beneficial for SC task.  

\subsection{Analysis of visual-based deep learning frameworks for scene classification}
\label{vi_res}
\begin{figure*}[t]
    \centering
    \includegraphics[width =1.0\linewidth]{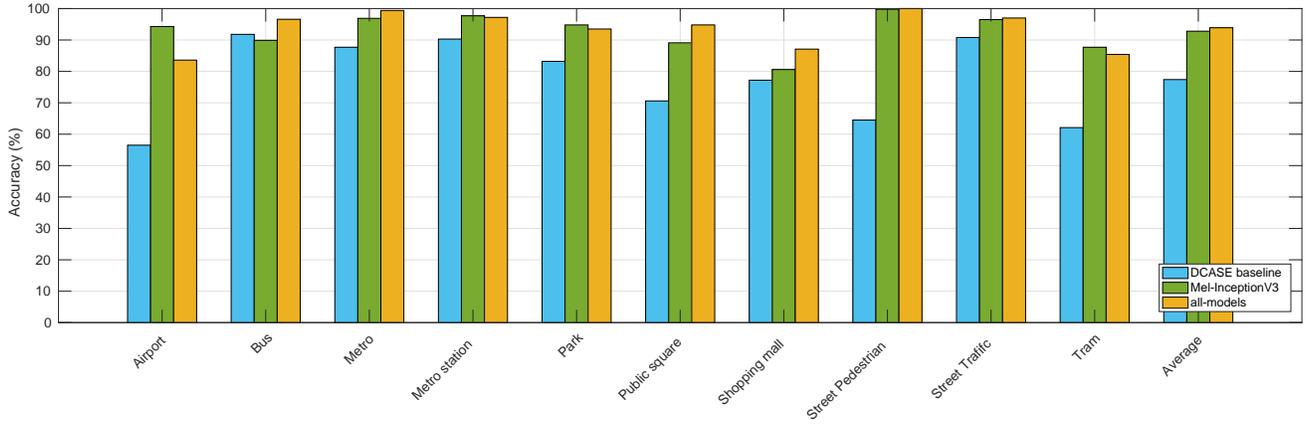}
    	\vspace{-0.5cm}
	\caption{Performance comparison (Acc.\%) of DCASE baseline, \textit{MEL-InceptionV3} and \textit{all-models} across all scene categories}
    \label{fig:F5}
\end{figure*}
%
\begin{figure}[t]
    \centering
    \includegraphics[width =1.0\linewidth]{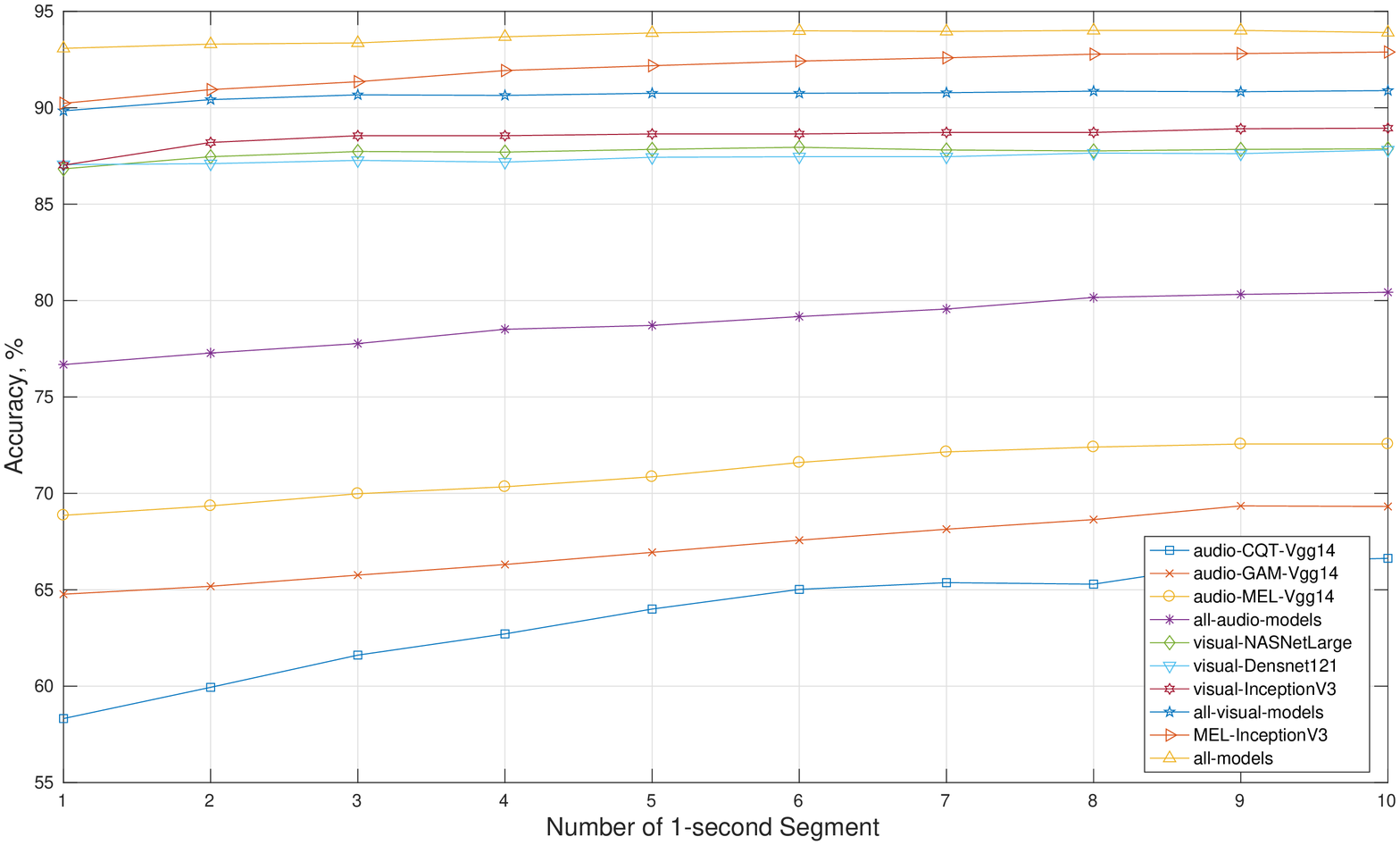}
    	\vspace{-0.5cm}
	\caption{Performance of individual audio-spectrogram based frameworks (\textit{audio-CQT-Vgg14, audio-GAM-Vgg14, audio-MEL-Vgg14}), PROD fusion of three audio-spectrogram based frameworks (\textit{all-audio-models}), individual visual-image based frameworks (\textit{visual-image-NASNetLarnge, visual-image-Densnet121, visual-image-InceptionV3}), PROD fusion of three visual-image based frameworks (\textit{all-visual-model}), PROD fusion of audio-based and visual based frameworks (\textit{MEL-InceptionV3, all-models}) with the increasing number of 1-second input segments}
    \label{fig:F6}
    \vspace{-0.6cm}
\end{figure}
As obtained results are shown in Table~\ref{table:vi_res}, we can see that the visual-image based frameworks, which directly train image frame inputs, outperform visual-embedding based frameworks.
While all late fusion methods over visual-image based frameworks help to improve the performance, only MAX fusion of image-embedding based frameworks shows to be effective.
The PROD fusion of seven visual-image based frameworks achieves the best accuracy of 91.1\%, improving DCASE baseline by 13.7\% (Note that these frameworks and DCASE baseline only use visual data input).
Comparing the performance between audio-based and visual-based approaches, the PROD fusion of seven visual-image based frameworks (91.1\%) outperforms the best result of 82.2\% from PROD fusion of \textit{audio-CQT-Vgg14, audio-GAM-Vgg14, audio-MEL-Vgg14,} and \textit{audio-emb-CNN14} mentioned in Section~\ref{au_res}.
Further comparing the two confusion matrixes obtained from these two PROD fusions, as shown in Fig.~\ref{fig:F1} and Fig.~\ref{fig:F2}, we can see that PROD fusion of seven visual-image based frameworks outperforms over almost scene categories except to 'Park' and 'Tram'.
As a result, we can conclude that visual data input contains more information for scene classification rather than audio data input.

\subsection{Combine both visual and audio features for scene classification}
As individual analysis of either audio or visual features within scene context classification is shown in Section~\ref{au_res} and~\ref{vi_res} respectively, we can see that directly training and classifying audio/visual data input is more effective, rather than audio/image-embedding based approaches. 
We then evaluate a combination of audio and visual features by proposing two PROD fusions: (1) three audio-spectrogram based frameworks (\textit{audio-CQT-Vgg14, audio-GAM-Vgg14, audio-MEL-Vgg14}) and top-3 visual-image frameworks (\textit{visual-image-DenseNet121, visual-image-InceptionV3, visual-image-NASNetLarge}) referred to as \textit{all-models}, and (2) one audio-spectroram based framework (\textit{audio-MEL-Vgg14}) and one visual-image based framework (\textit{visual-image-InceptionV3}) referred to as \textit{MEL-InceptionV3}.
As results shown in Fig.~\ref{fig:F5}, \textit{all-models} helps to achieve the highest accuracy classification score of 93.9\%, improving DCASE baseline by 16.5\% and showing improvement on all scene categories.
Although \textit{MEL-InceptionV3} only fuses two frameworks, it achieves 92.8\%, showing competitive to \textit{all-models} fusing 6 frameworks.
Notably, mis-classifying cases mainly occur among high cross-correlated categories in meta-class such as indoor (Airport, Metro station, and Shopping mall), outdoor (Park, Public square, Street pedestrian, and Street Traffic), and transportation (Bus, Metro, and Tram).
If we aim to classify into three meta-classes (indoor, outdoor, and transportation), \textit{all-models} helps to achieve a classification accuracy of 99.3\%.

\subsection{Early detecting scene context}
We further evaluate whether deep learning frameworks can help to detect scene context early.
To this end, we evaluate 10 different frameworks: (1-2-3) 3 individual audio-spectrogram based frameworks (\textit{audio-CQT-Vgg14, audio-GAM-Vgg14, audio-MEL-Vgg14}), (4) PROD fusion of these three audio-spectrogram frameworks referred to as \textit{all-audio-models}, (5-6-7) 3 visual-image based frameworks (\textit{visiual-image-NASNetLarge, visual-image-Densnet121, visual-image-InceptionV3}), (8) PROD fusion of these three visual-image based frameworks referred to as \textit{all-visual-models}, (9) \textit{MEL-InceptionV3}, and (10) \textit{all-models}.
As the results shown in Fig.~\ref{fig:F6}, while performance of audio-based frameworks is improved by time, visual-based frameworks are stable.
As a result, when we combine audio and visual features, which are evaluated in \textit{MEL-InceptionV3} and \textit{all-models}, the performance is improved by time and stable after 6 seconds. 
Notably, accuracy scores of both \textit{MEL-InceptionV3} and \textit{all-models} are larger than 90.0\% at the first second, which is potentially for real-life applications integrating the function of early detecting scene context.

\section{Conclusion}
We conducted extensive experiments and explored various deep learning based frameworks for classifying 10 categories of urban scenes.
Our method, which uses an ensemble of audio-based and visual-based frameworks, achieves the best classification accuracy of 93.9\% on DCASE Task 1B development set.
The obtained results outperform DCASE baseline, improving by 17.1\% with only audio data input, 26.2\% with only visual data input, and 16.5\% with both audio-visual data.
In further work, we will evaluate whether an end-to-end system using joint learning of audio-visual data input may help to improve the performance.

\addtolength{\textheight}{-12cm}   

\bibliographystyle{IEEEbib}
\bibliography{refs}

\end{document}